\documentclass[conference]{IEEEtran}
\IEEEoverridecommandlockouts

\usepackage{cite}
\usepackage{amsmath,amssymb,amsfonts}
\usepackage{graphicx}
\usepackage{textcomp}
\usepackage[dvipsnames]{xcolor}
\usepackage{multirow}
\usepackage{booktabs}
\usepackage{enumitem}
\usepackage{mathtools}

\usepackage{framed}
\usepackage{colortbl}

\usepackage{graphicx}
\definecolor{B}    {HTML}{000000}
\definecolor{B2}   {HTML}{003399}
\definecolor{Bv}   {HTML}{0000EB}
\definecolor{R}    {HTML}{c9171e}
\definecolor{R2}   {HTML}{d7003a}
\definecolor{INK}  {HTML}{595857}
\definecolor{Y}    {HTML}{f1c40f}
\definecolor{G}    {HTML}{009a00}
\definecolor{GRAY} {HTML}{808080}
\definecolor{MAUVE}{HTML}{9400D1}
\definecolor{crimson}{rgb}{0.86, 0.08, 0.24}
\definecolor{royalblue}{RGB}{65, 105, 225}

\usepackage{amsmath,amsfonts}
\usepackage{mathtools}
\usepackage{bm}
\usepackage{amsthm}
\usepackage{pifont}%
\newcommand{\cmark}{\ding{51}}%
\newcommand{\xmark}{\ding{55}}%

\usepackage{array}
\usepackage{booktabs}
\usepackage{colortbl}
\usepackage{multirow}
\usepackage{tabu}

\usepackage{url, hyperref}
\hypersetup{
    colorlinks, linkcolor=royalblue, citecolor=royalblue, urlcolor=royalblue,
    allcolors=royalblue,
    breaklinks=true,
    pdfborder={0 0 0}}

\usepackage{enumitem}
\usepackage{lipsum}
\usepackage[plain]{algorithm}
\floatstyle{plaintop}
\usepackage{algpseudocode}
\algrenewcommand{\alglinenumber}[1]{{\scriptsize\bfseries\ttfamily\color{R}#1}}
\floatstyle{plaintop}
\restylefloat{algorithm}

\usepackage{xpatch}%
\makeatletter
\xpatchcmd{\algorithmic}{\ALG@tlm\z@}{\ALG@tlm\z@\leftmargin 10pt}{}{}
\makeatother

\usepackage{listings}
\usepackage{caption, subcaption}
\usepackage{setspace}

\usepackage{soul}%

\colorlet{shadecolor}{royalblue!15}
\usepackage{framed}

\newenvironment{columnshaded}{
  \setlength{\fboxsep}{6pt}%
  \begin{center}
  \begin{minipage}{\dimexpr\columnwidth - 2\fboxsep\relax}
  \setlength{\parindent}{0pt}%
  \begin{snugshade}
}{
  \end{snugshade}
  \end{minipage}
  \end{center}
}

\ifx\UseACMTemplate\undefined
\else

\fi

\ifx\UseIEEETemplate\undefined
\else

\fi

\usepackage{comment}

\usepackage{tikz}
\usetikzlibrary{positioning}
\usetikzlibrary{shapes.multipart}
\usetikzlibrary{shapes.geometric}
\usetikzlibrary{arrows}
\usetikzlibrary{arrows.meta}
\usetikzlibrary{decorations.text}
\usetikzlibrary{decorations.pathreplacing}
\usepgflibrary{decorations.markings}
\usetikzlibrary{backgrounds}
\usetikzlibrary{patterns}
\usetikzlibrary{fit}
\usetikzlibrary{shadows.blur}
\usetikzlibrary{shapes.symbols}

\usepackage{main}

\definecolor{shadecolor}{rgb}{0.89,0.91,0.98}

\definecolor{CornflowerBlue}{RGB}{100,149,237}
\newcommand{\step}[1]{\tikz[baseline=(char.base)]{%
  \node[shape=circle,fill=RoyalBlue,text=white,inner sep=0.5pt,outer sep=0pt,
        font=\footnotesize\bfseries,minimum size=1.1em] (char) {#1};}}
\newcommand{\dstep}[1]{\tikz[baseline=(char.base)]{%
  \node[shape=circle,fill=CornflowerBlue,text=white,inner sep=0.5pt,outer sep=0pt,
        font=\footnotesize\bfseries,minimum size=1.1em] (char) {#1};}}

\colorlet{HLCOLOR}{B}
\colorlet{TableAltColor}{gray!20}

\newcommand{\FloatBodyStyle}{\centering\small\fontfamily{lmr}\selectfont}

\newcommand{\BestCr}[1]{\color{royalblue} \usefont{OT1}{cmr}{bx}{n} {#1}}

\newcommand{\fzgpu}{FZ-GPU}

\newcommand{\cuszp}{\mbox{cuSZp}}
\newcommand{\cuszpP}{\mbox{cuSZp-P}}
\newcommand{\cuszpO}{\mbox{cuSZp-O}}
\newcommand{\fsz}{\mbox{FSZ}}

\makeatletter
\long\def\aptLtoX{\@ifnextchar[{\@aptLtoX}{\@aptLtoX[]}}
\long\def\@aptLtoX[#1]#2#3{#3}
\makeatother

\begin{document}

\title{{\fsz}: Breaking the Prediction-Throughput Trade-off in GPU Lossy Compression}

\author{
\IEEEauthorblockN{Jiajun Huang}
\IEEEauthorblockA{University of South Florida, Tampa, FL, USA \\
jiajunhuang@usf.edu}
}

\maketitle
\thispagestyle{plain}

\begin{abstract}
Existing fast GPU error-bounded lossy compressors have achieved high throughput through pure-GPU single-kernel designs, but their compression ratios remain limited because they typically apply a fixed first-order predictor on independent blocks. We propose {\fsz}, a GPU error-bounded lossy compressor that redesigns the prediction stage with three mutually reinforcing algorithmic innovations to achieve both higher compression ratios and higher throughput within a single CUDA kernel: (1) \emph{cross-block prediction state} carries Lorenzo prediction state across block boundaries within 256-element tiles, eliminating 7 out of 8 boundary residuals that inflate encoding rates; (2) \emph{per-tile adaptive multi-order prediction and centering} adaptively selects the best compression strategy per tile from first-order, second-order, and centering variants; and (3) a \emph{single-pass four-way evaluation} exploits a mathematical property of finite differences to evaluate all variants from a single data read, enabling richer prediction within the same bandwidth budget as a fixed predictor. Experiments on NVIDIA GH200 GPU with 8 real-world application datasets show that {\fsz} outperforms {\cuszpP} by up to 10.95$\times$ and the state-of-the-art {\cuszpO} by up to 2.92$\times$ in compression ratio. Notably, these gains come with no throughput penalty: {\fsz} simultaneously achieves the highest average throughput (676~GB/s compression, 785~GB/s decompression) among all evaluated compressors.\end{abstract}


\section{Introduction}
\label{sec:intro}

GPU-accelerated scientific and AI applications increasingly demand \emph{inline} compression at hundreds of GB/s. Free-electron laser facilities such as LCLS generate X-ray data at approximately 250~GB/s~\cite{Huang2024cuSZp2}, large-scale GPU simulations in cosmology, climate, and turbulence produce massive volumes of floating-point data~\cite{Zhao2020SDRBench}, and AI training increasingly relies on compression to reduce communication and storage costs~\cite{Huang2024cuSZp2, Huang2025ZCCL, huang2023ccoll, huang2023gzccl, Huang2025ghZCCL}. Error-bounded lossy compression addresses this by introducing controlled distortion within a user-specified threshold, achieving compression ratios of 10--100$\times$ while preserving data fidelity~\cite{Di2016SZ, Tao2017SZ, Lindstrom2014ZFP}. To avoid becoming a bottleneck in these GPU-centric workflows, the compressor must operate at hundreds of GB/s end-to-end.

\looseness=-1
Recent GPU lossy compressors such as {\cuszp}~\cite{cuSZp2023, Huang2024cuSZp2} have pushed throughput to hundreds of GB/s by fusing the entire pipeline into a single GPU kernel. However, these advances have focused primarily on \emph{throughput engineering}: optimizing memory access patterns, encoding strategies, and synchronization latency. The \emph{prediction} stage, the step that decorrelates the data and ultimately determines how many bits each element needs, has been kept deliberately simple to avoid throughput loss. State-of-the-art pure-GPU compressors typically use a fixed first-order predictor on independent blocks, accepting the resulting compression ratio limitations as the price of GPU parallelism.

\subsection{Limitations of Existing Works}

We identify three key challenges in the prediction strategies of state-of-the-art pure-GPU error-bounded lossy compressors, as summarized in Table~\ref{tab:compressor_comparison}.

\begin{table}[ht]
    \centering
    \FloatBodyStyle\footnotesize
    \renewcommand{\arraystretch}{1.15}
    \resizebox{\columnwidth}{!}{
    \begin{tabular}{l ccc ccc}
        \toprule
        & \multicolumn{3}{c}{\textbf{Compression Ratio}} & \multicolumn{3}{c}{\textbf{Throughput}} \\
        \cmidrule(lr){2-4} \cmidrule(lr){5-7}
        \textbf{\shortstack[l]{GPU Lossy \\ Compressor}} &
        \textbf{\shortstack{Cross-Blk \\ State}} &
        \textbf{\shortstack{Multi-Order \\ Prediction}} &
        \textbf{\shortstack{Adaptive \\ Centering}} &
        \textbf{\shortstack{Single-Pass \\ Evaluation}} &
        \textbf{\shortstack{Single \\ Kernel}} &
        \textbf{\shortstack{Pure \\ GPU}} \\
        \midrule
        \rowcolor{gray!10} {\cuszpP}~\cite{cuSZp2023} & \textcolor{red}{\textbf{\xmark}} & \textcolor{red}{\textbf{\xmark}} & \textcolor{red}{\textbf{\xmark}} & \textcolor{red}{\textbf{\xmark}} & \textcolor{RoyalBlue}{\textbf{\cmark}} & \textcolor{RoyalBlue}{\textbf{\cmark}} \\
        {\cuszpO}~\cite{Huang2024cuSZp2} & \textcolor{red}{\textbf{\xmark}} & \textcolor{red}{\textbf{\xmark}} & \textcolor{red}{\textbf{\xmark}} & \textcolor{red}{\textbf{\xmark}} & \textcolor{RoyalBlue}{\textbf{\cmark}} & \textcolor{RoyalBlue}{\textbf{\cmark}} \\
        \rowcolor{gray!10} {\fzgpu}~\cite{ZhangFZ-GPU2023} & \textcolor{red}{\textbf{\xmark}} & \textcolor{red}{\textbf{\xmark}} & \textcolor{red}{\textbf{\xmark}} & \textcolor{red}{\textbf{\xmark}} & \textcolor{red}{\textbf{\xmark}} & \textcolor{RoyalBlue}{\textbf{\cmark}} \\
        \textbf{{\fsz} (ours)} & \textcolor{RoyalBlue}{\textbf{\cmark}} & \textcolor{RoyalBlue}{\textbf{\cmark}} & \textcolor{RoyalBlue}{\textbf{\cmark}} & \textcolor{RoyalBlue}{\textbf{\cmark}} & \textcolor{RoyalBlue}{\textbf{\cmark}} & \textcolor{RoyalBlue}{\textbf{\cmark}} \\
        \bottomrule
    \end{tabular}
    }
    \vspace{3pt}
    \caption{Comparison of key designs in pure-GPU error-bounded lossy compressors. {\fsz} is the only compressor with all compression ratio and throughput features.}
    \label{tab:compressor_comparison}
\end{table}

\textbf{(C1) Independent block processing amplifies boundary overhead.} To enable GPU parallelism, existing fast GPU lossy compressors partition data into small independent blocks (typically 32 elements) and restart prediction state at every boundary. The first element of each block has no predecessor, producing a raw quantized value as its ``residual,'' often orders of magnitude larger than the intra-block differences. Because fixed-length encoding sets the block's bit rate from its \emph{maximum} residual, a single large boundary value inflates the rate for all 32 elements in the block. This boundary overhead grows with the number of blocks and is amplified across millions of blocks in a typical scientific dataset; yet resolving it requires maintaining state across blocks, which introduces sequential dependencies that conflict with GPU parallelism.

\looseness=-1
\textbf{(C2) Fixed prediction cannot adapt to diverse data patterns.} Existing pure-GPU compressors apply a fixed first-order predictor (first-difference) uniformly across the entire dataset. Yet scientific data exhibits diverse local characteristics: some fields are piecewise constant, others follow smooth polynomial trends, and others carry large constant offsets (e.g., atmospheric pressure at 1000\,hPa) that inflate the first element's residual. A single fixed predictor cannot be optimal for all these patterns, producing unnecessarily large residuals on a significant fraction of tiles. On a GPU, adapting the prediction strategy is non-trivial: evaluating multiple variants requires additional data passes that consume scarce memory bandwidth, and per-tile adaptive decisions risk introducing warp divergence.

\textbf{(C3) Richer prediction must not sacrifice throughput.} Fast GPU lossy compressors are memory-bandwidth-bound: on modern GPUs, even a single extra global memory read per element can reduce throughput by 30--50\%. Naively evaluating multiple prediction strategies would require multiple passes over the input data, multiplying bandwidth consumption proportionally. Any solution to C1 and C2 must therefore fit within the same bandwidth and compute budget as the existing fixed predictor, or the throughput gains of single-kernel compression would be lost.

\subsection{Our Solution: {\fsz}}

To address these challenges, we propose \textbf{{\fsz}}, a GPU error-bounded lossy compressor that achieves both high compression ratio and high throughput within a single CUDA kernel. To the best of our knowledge, {\fsz} is the first pure-GPU single-kernel compressor with cross-block prediction state and adaptive multi-order prediction. The three designs form a virtuous cycle: each unlocks the full potential of the others.

\begin{itemize}[leftmargin=10pt]

\item \textbf{Cross-block prediction state} creates a continuous 256-element prediction chain by carrying prediction state across 8 block boundaries via warp registers, reducing boundary residuals from 8 per tile to just 1. Since a single thread processes all 8 blocks of its tile sequentially, this continuity is achieved with zero synchronization cost while inter-tile parallelism is fully preserved.

\item \textbf{Per-tile adaptive multi-order prediction and centering} selects the best compression strategy per tile from four variants (first-order and second-order Lorenzo prediction, each with and without adaptive centering). These techniques are mutually reinforcing: cross-block state eliminates boundary overhead, adaptive centering removes constant offsets, and second-order prediction captures linear trends, each targeting a different source of compression ratio gain.

\item \textbf{Single-pass four-way evaluation} enables all four variants to be evaluated from a \emph{single} data read by exploiting the mathematical property that constant offsets cancel exactly in $k$-th order finite differences. Combined with vectorized \texttt{float4} memory access, PTX-level quantization, and decoupled lookback prefix-sum~\cite{MerrillDecoupledLookback2016} within a single fused kernel, {\fsz} achieves \emph{higher} throughput than existing fast GPU lossy compressors while performing $4\times$ richer prediction.

\item \textbf{Comprehensive evaluation} on NVIDIA GH200 with 8 real-world scientific datasets (84 fields, 3 error bounds) shows that {\fsz} outperforms the state-of-the-art {\cuszpO} by up to 2.92$\times$ and {\cuszpP} by up to 10.95$\times$ in compression ratio, while simultaneously achieving the highest throughput (676~GB/s compression, 785~GB/s decompression) among all evaluated compressors.

\end{itemize}

We first challenge the assumption that richer prediction inevitably slows down GPU compressors by showing that the mandatory two-pass architecture leaves arithmetic headroom that can be exploited for richer prediction (Section~\ref{sec:rethinking}), then present {\fsz}'s design (Sections~\ref{sec:high_level}--\ref{sec:detailed_design}) and evaluate it on 8 scientific datasets (Sections~\ref{sec:evaluation}--\ref{sec:eval-further}). Section~\ref{sec:related} discusses related work and Section~\ref{sec:conclusion} concludes.

\section{Rethinking Prediction in Fast GPU Lossy Compressors}
\label{sec:rethinking}

Existing fast GPU lossy compressors have achieved extreme throughput by optimizing encoding strategies, memory access patterns, and synchronization mechanisms~\cite{cuSZp2023, Huang2024cuSZp2, ZhangFZ-GPU2023}. However, they typically use a fixed first-order predictor on independent blocks. This design choice is motivated by a widely held assumption: \emph{richer prediction inevitably slows down GPU compressors due to additional data accesses and computation.} In this section, we challenge this assumption.

\subsection{The Two-Pass Constraint}

Single-kernel variable-length encoding requires two passes over the input data:
\begin{enumerate}[leftmargin=15pt]
\item \textbf{Rate pass:} Read data, quantize, predict, and determine per-block rates (bits per element).
\item \textbf{Encode pass:} Read data again, apply prediction, and write encoded output to byte offsets computed by a global prefix-sum on the rates.
\end{enumerate}
The encode pass cannot begin until the prefix-sum completes, which depends on rates from the first pass. This two-pass constraint is \emph{fundamental}: it cannot be eliminated without either abandoning variable-length encoding (sacrificing CR) or abandoning single-kernel execution (sacrificing throughput). Existing single-kernel compressors such as {\cuszp}~\cite{cuSZp2023, Huang2024cuSZp2, Huang2025cuSZp3} face the same dependency, buffering quantized values in thread-local memory across the prefix-sum.

\subsection{The Bandwidth-Compute Gap}

The key observation is that the rate pass is \emph{memory-bandwidth-bound}, not compute-bound. On NVIDIA GH200 (4~TB/s HBM3 bandwidth), reading a 1\,KB data block takes approximately the same time as hundreds of integer arithmetic operations. A single-variant compressor (e.g., {\cuszpO} with LZ1-only) performs a quantization, one Lorenzo subtraction, sign extraction, and a max comparison per element, consuming a small fraction of the arithmetic throughput.

This leaves \textbf{arithmetic headroom} in the rate pass: the ALU is partially idle while waiting for memory. In principle, additional prediction logic can be inserted into the rate pass without slowing it down, as long as it does not introduce extra global memory reads or excessive register pressure. The question is whether this headroom can be exploited to meaningfully improve compression ratio.

\subsection{From Constraint to Opportunity}

The existence of arithmetic headroom is necessary but not sufficient. Exploiting it to improve compression ratio faces a fundamental tension: the very techniques that improve prediction quality, such as longer-range dependencies, richer models, and data-adaptive decisions, are precisely those that conflict with GPU execution constraints, including lock-step warps, limited registers, and coalesced memory access patterns. Any improvement to the prediction stage must fit within the rate pass without introducing extra global memory reads, thread synchronization, or warp divergence. Prior fast GPU compressors have treated this tension as irreconcilable.

{\fsz} resolves this tension. Rather than treating prediction quality and GPU efficiency as opposing goals, {\fsz} co-designs its prediction stage with the GPU execution model so that richer prediction aligns naturally with warp-level execution, register allocation, and memory access patterns. The following sections show how this co-design simultaneously improves both compression ratio and throughput.

\section{{\fsz} High-Level Design}
\label{sec:high_level}

\subsection{{\fsz} Overview}

{\fsz} is a single-kernel GPU error-bounded lossy compressor for single-precision floating-point scientific data. The entire compression and decompression pipelines are each executed within a single CUDA kernel, with no CPU-GPU round-trips, no multi-kernel orchestration, and no host-side computation.

\textbf{Key idea: three complementary designs within a tile framework.} Existing fast GPU compressors typically process data in independent blocks (commonly 32 elements), using a fixed first-order predictor. {\fsz} introduces three mutually reinforcing designs: (1)~\emph{cross-block prediction state} groups 8 blocks into a 256-element tile and carries prediction state across boundaries, eliminating 7 out of 8 boundary residuals; (2)~\emph{adaptive multi-order prediction and centering} selects per tile among four prediction variants (LZ1/LZ2 $\times$ with/without centering), capturing diverse data patterns and reducing constant offsets; and (3)~a \emph{single-pass four-way evaluation} that exploits a mathematical property of finite differences to evaluate all four variants from a single data read, enabling richer prediction without sacrificing throughput.

\textbf{Data organization.} Each tile consists of 8 blocks of 32 elements. The block size of 32 packs each bit-plane of a block into exactly one 32-bit word and the sign map into 4 bytes, so every encoded write is a 4-byte-aligned \texttt{uchar4} store. Each CUDA thread processes 4 tiles (1024 elements) sequentially, and each CUDA thread block is a single warp (32 threads). All threads in a warp process neighboring tiles, ensuring coalesced memory access and efficient intra-warp communication via shuffle operations. A sensitivity analysis of tile configurations is presented in Section~\ref{sec:eval-tilesize}.


\begin{figure}[t]
\centering
\includegraphics[width=\columnwidth]{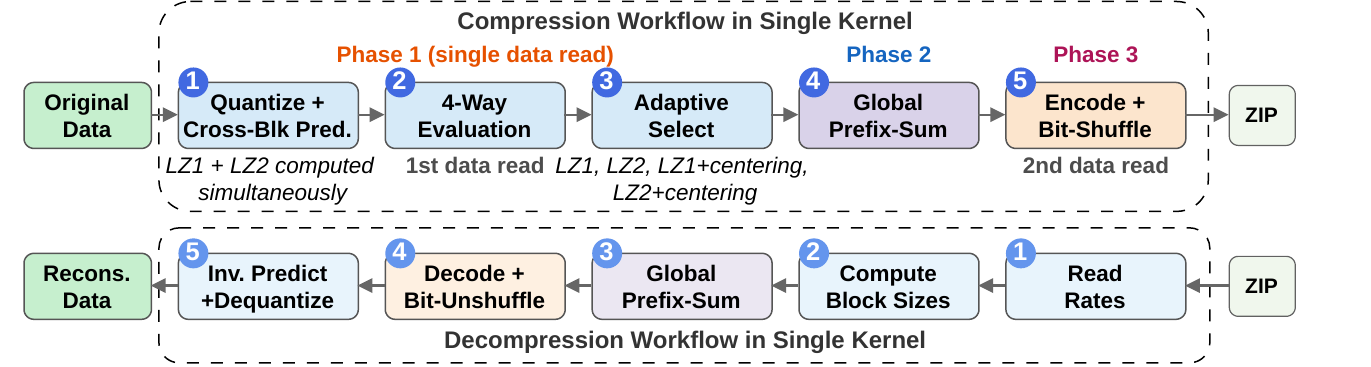}
\caption{Overview of the {\fsz} compression (top) and decompression (bottom) workflows, both executing within a \emph{single} CUDA kernel. Steps \protect\step{1}--\protect\step{3} are fused into a single data read (Phase~1): quantize with cross-block prediction, evaluate all four prediction variants, and adaptively select the best. Step \protect\step{4} computes global byte offsets via decoupled lookback (Phase~2). Step \protect\step{5} re-reads data and encodes with the selected variant (Phase~3).}
\label{fig:fsz_pipeline}
\end{figure}

\textbf{Compression pipeline.} We use Figure~\ref{fig:fsz_pipeline} to illustrate the workflow of {\fsz} compression. The two-pass structure is inherent to single-kernel variable-length encoding: the first read computes per-block rates, a global prefix-sum converts rates to byte offsets, and the second read encodes to those offsets. Even a single-variant compressor (e.g., LZ1-only) requires this two-pass design. {\fsz} exploits the first read's spare computational capacity to evaluate richer prediction at near-zero extra cost: maintaining LZ2 state requires only two additional registers per thread, and accumulating the tile sum for centering adds one floating-point addition per element.

\looseness=-1
The compression kernel consists of five steps. First, \step{1}~\textbf{Quantize + Cross-Block Predict} reads the input data once via vectorized \texttt{float4} loads, quantizes every floating-point value to an integer (the only lossy step), and simultaneously computes both LZ1 and LZ2 residuals with cross-block prediction state carried across all 8 blocks within each tile. Second, \step{2}~\textbf{Single-Pass 4-Way Evaluation} evaluates the encoding cost of four prediction variants (LZ1, LZ2, LZ1+centering, LZ2+centering) in the same pass as \step{1}, requiring no additional data read. Third, \step{3}~\textbf{Adaptive Select} picks the variant with the lowest cost per tile and writes the rate header. Steps \step{1}--\step{3} are fused into a single data read (Phase~1). Fourth, \step{4}~\textbf{Global Prefix-Sum} computes byte offsets for the compressed output using a three-level hierarchical prefix-sum: thread-level, warp-level via \texttt{\_\_shfl\_up\_sync}, and device-level via decoupled lookback~\cite{MerrillDecoupledLookback2016} (Phase~2). Finally, \step{5}~\textbf{Encode + Bit-Shuffle} re-reads the input data, applies only the selected prediction variant with cross-block state, and encodes residuals using fixed-length bit-shuffled encoding with coalesced \texttt{uchar4} writes (Phase~3). Rate-0 blocks (all residuals zero) store nothing and update the predictor state analytically, avoiding element-by-element computation.

\textbf{Compressed format.} The output consists of a rate header (one byte per block) followed by variable-length encoded data:
\begin{itemize}[leftmargin=10pt]
\item \textbf{Block 0 of each tile:} Bit 7 = LZ order, bit 6 = centering flag, bits 4--0 = rate (0--31).
\item \textbf{Blocks 1--7:} Bits 4--0 = rate.
\item \textbf{Encoded data per block:} 4-byte sign map + $\textit{rate} \times 4$ bytes of bit-shuffled residuals. Zero-rate blocks store nothing.
\item \textbf{Tile mean (optional):} 4-byte float stored inline when centering is enabled.
\end{itemize}

\textbf{Decompression.} As shown in the bottom row of Figure~\ref{fig:fsz_pipeline}, {\fsz} decompression reverses the pipeline within a single CUDA kernel: \dstep{1}~\textbf{Read Rates} reads the rate header and tile metadata (LZ order, centering flag); \dstep{2}~\textbf{Compute Block Sizes} calculates the compressed size of each block from the rates; \dstep{3}~\textbf{Global Prefix-Sum} computes byte offsets via decoupled lookback; \dstep{4}~\textbf{Decode + Bit-Unshuffle} reconstructs residuals from the compressed stream; and \dstep{5}~\textbf{Inv.\ Predict + Dequantize} applies inverse cross-block prediction with the selected LZ order and converts quantized integers back to floating-point values. Rate-0 blocks are reconstructed directly in the float domain: LZ1 produces a constant fill, while LZ2 produces an arithmetic sequence, both computed in closed form without decoding. When the compression ratio exceeds 100$\times$, an adaptive pre-zero optimization skips writes for zero-rate blocks entirely.

\section{Core Design of {\fsz}}
\label{sec:detailed_design}

This section presents {\fsz}'s three novel designs, each targeting a different challenge from Section~\ref{sec:intro}: cross-block prediction state (Section~\ref{sec:cross_block}, addressing C1) eliminates boundary overhead, adaptive multi-order prediction with centering (Section~\ref{sec:adaptive_pred}, addressing C2) captures diverse data patterns, and single-pass four-way evaluation (Section~\ref{sec:four_way_eval}, addressing C3) enables both without additional bandwidth. The three designs reinforce each other: cross-block state makes interior residuals dominant, which adaptive prediction then drives to zero. The encoding and synchronization mechanisms (Sections~\ref{sec:encoding}--\ref{sec:sync}) build on established techniques~\cite{Huang2024cuSZp2,MerrillDecoupledLookback2016} optimized for {\fsz}'s tile architecture.

\subsection{Cross-Block Prediction State}
\label{sec:cross_block}

\begin{figure}[t]
\centering
\includegraphics[width=\columnwidth]{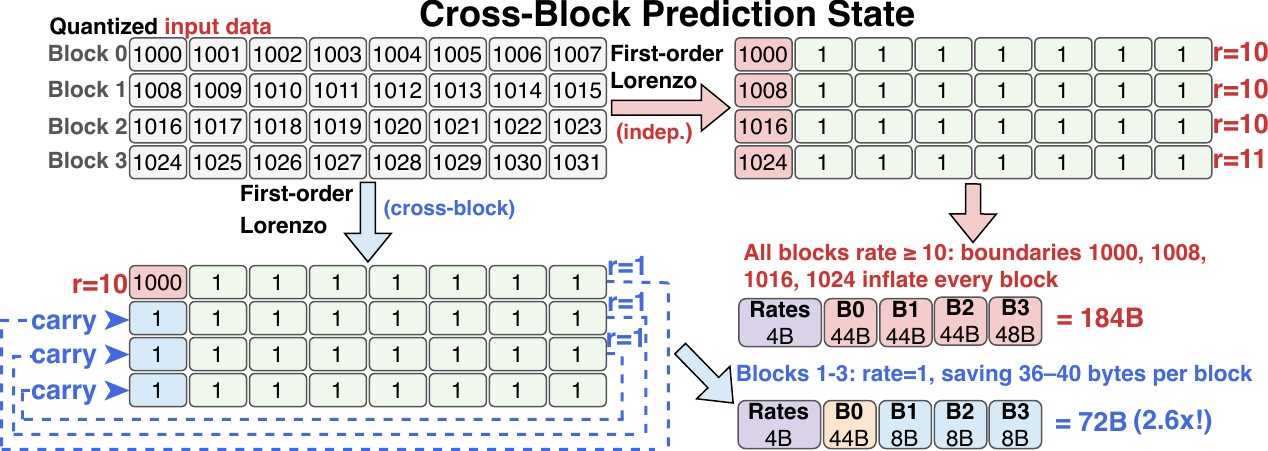}
\caption{Cross-block prediction state. (a)~Independent blocks: each boundary is a raw value (1000, 1008, 1024), inflating every element's rate to 10--11. (b)~Cross-block: {\fsz} carries the last value across boundaries, cutting the rate to 1 and saving up to 40 bytes per block.}
\label{fig:cross_block}
\end{figure}

\subsubsection{Limitation of Independent Blocks}

As illustrated in Figure~\ref{fig:cross_block}, in a standard block-based GPU compressor, prediction state is restarted at every 32-element block boundary. The first element of each block has no predecessor, so its ``residual'' is the raw quantized value, far larger than the intra-block differences. Within a 256-element tile (8 blocks), 8 elements produce large residuals that inflate the per-block rate.

At tight error bounds, where the typical residual magnitude is small (e.g., 2--8 bits), a single large boundary value can double the block's rate, wasting 31 elements worth of encoding capacity. This overhead is amplified across millions of blocks in a typical scientific dataset.

\subsubsection{Continuous Prediction Chain}

{\fsz} maintains prediction state across all 8 block boundaries within each tile. Let block $b$ contain elements $q_{32b}, q_{32b+1}, \ldots, q_{32b+31}$. In standard independent-block processing, the LZ1 residual for the first element of each block is:
\begin{equation}
\ell_{32b}^{(1)} = q_{32b} \quad \text{(no predecessor, raw value)}
\label{eq:indep_boundary}
\end{equation}
With cross-block state, {\fsz} instead computes:
\begin{equation}
\ell_{32b}^{(1)} = q_{32b} - q_{32(b-1)+31} = q_{32b} - q_{32b-1}
\label{eq:crossblock_boundary}
\end{equation}
using the last element of the previous block as the predecessor. For LZ2, the last two values ($q_{32b-2}$, $q_{32b-1}$) carry forward.

This creates a continuous 256-element prediction chain. Only the very first element of the tile ($q_0$) lacks a predecessor, reducing boundary overhead from 8 large residuals per tile to 1. Since the rate of a block is determined by $r_b = \lceil \log_2\bigl(\max_j |\ell_j| + 1\bigr) \rceil$, eliminating a single large $|\ell_{32b}|$ from each block can significantly reduce $r_b$ and thus the block's compressed size.

\textbf{Rate-0 block state update.} When $r_b = 0$ (all residuals are zero), the predictor state can be updated analytically: for LZ1, $\ell_i = 0$ implies $q_i = q_{32b}$ for all $i$ in the block, so the carry-forward value is $q_{32b+31} = q_{32b}$. For LZ2, $\ell_i = 0$ implies the values form an arithmetic sequence $q_{32b+j} = q_{32b} + j \cdot s$ where $s = q_{32b} - q_{32b-1}$, enabling closed-form state advancement without element-by-element computation.

\textbf{Parallelism trade-off.} Cross-block state introduces a sequential dependency within each tile: block $b{+}1$'s boundary residual depends on block $b$'s last value. On GPUs, where parallelism drives performance, such dependencies are typically avoided. In {\fsz}, a single thread processes all 8 blocks of its tile in a sequential loop, so the dependency is satisfied naturally in registers. Inter-tile parallelism is fully preserved.

\aptLtoX[graphic=no,type=html]{\begin{shaded}\textbf{Takeaway I}: Cross-block prediction imposes a sequential dependency within each tile. {\fsz} resolves this at zero cost by aligning the dependency with the thread model: one thread processes all 8 blocks sequentially, so inter-block state is simply carried in registers with no synchronization.\end{shaded}}{\begin{columnshaded}
\vspace{3pt}
\textbf{Takeaway I}: Cross-block prediction imposes a sequential dependency within each tile. {\fsz} resolves this at zero cost by aligning the dependency with the thread model: one thread processes all 8 blocks sequentially, so inter-block state is simply carried in registers with no synchronization.
\end{columnshaded}}

\subsection{Adaptive Multi-Order Prediction and Centering}
\label{sec:adaptive_pred}

Cross-block prediction state improves compression ratio by creating continuous prediction chains. To further capture diverse data patterns, {\fsz} introduces two per-tile adaptive techniques, multi-order prediction and centering, evaluated simultaneously via a single-pass four-way evaluation.

\begin{figure*}[t]
\centering
\begin{minipage}[t]{0.48\textwidth}
\centering
\includegraphics[width=\textwidth]{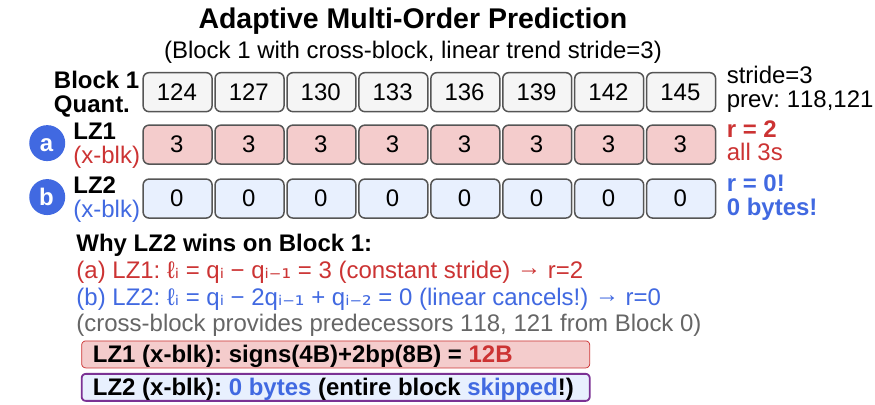}
\caption{Adaptive multi-order prediction on Block~1 with cross-block state (linear trend, stride=3). (a)~LZ1 residuals are constant~3 ($r{=}2$, cost 12\,B). (b)~LZ2 residuals are all zero ($r{=}0$) because $\Delta^2$ cancels linear trends; the block is skipped at zero cost.}
\label{fig:lz1_vs_lz2}
\end{minipage}
\hfill
\begin{minipage}[t]{0.48\textwidth}
\centering
\includegraphics[width=\textwidth]{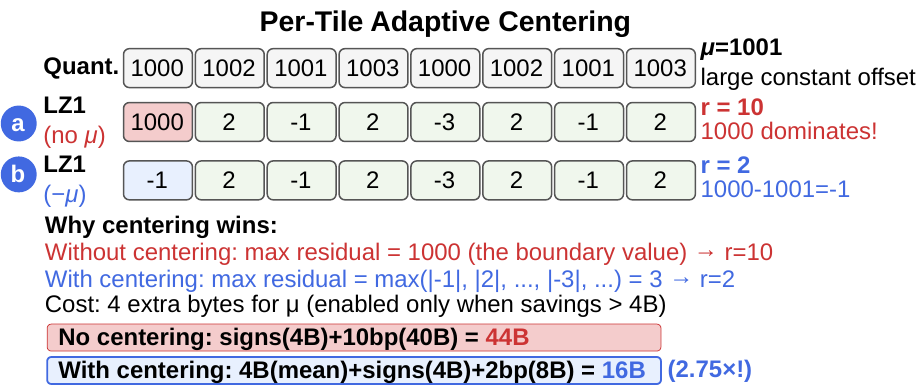}
\caption{Per-tile adaptive centering on data with a large constant offset ($\mu{=}1001$). (a)~Without centering, the boundary residual 1000 dominates, forcing $r{=}10$ (cost 44\,B). (b)~With centering, the boundary drops to $-1$, reducing $r$ to~2 (cost 16\,B including 4\,B for $\mu$), a 2.75$\times$ saving.}
\label{fig:centering}
\end{minipage}
\end{figure*}

\subsubsection{Why Adaptive Prediction Matters}

Scientific datasets exhibit fundamentally different local structures. Some fields are piecewise constant (cloud fraction, precipitation), where first-order differences are near-zero and LZ1 is optimal. Others follow smooth linear trends (temperature profiles, pressure gradients), where first-order differences remain large but second-order differences nearly vanish. A fixed first-order predictor cannot exploit these patterns, producing unnecessarily large residuals on a significant fraction of data.

Furthermore, cross-block state and adaptive prediction are mutually reinforcing: cross-block state reduces boundary residuals, making interior residuals the dominant factor in the block rate. As illustrated in Figure~\ref{fig:lz1_vs_lz2}, LZ1 still incurs a constant residual of~3 per element, requiring rate~2 (\step{a}), whereas LZ2 drives all residuals to zero on linear data, producing a rate-0 block that costs nothing (\step{b}). This benefit would not materialize without cross-block state providing predecessors from the previous block.

\subsubsection{Four-Way Cost Analysis}

Let $q_0, q_1, \ldots, q_{255}$ denote the quantized integers for a tile, where $q_i = \mathrm{round}(d_i / (2 \cdot eb))$. For each tile, {\fsz} evaluates four prediction variants:

\begin{enumerate}[leftmargin=15pt]
\item \textbf{LZ1}: $\ell_i^{(1)} = q_i - q_{i-1}$, \quad $i \ge 1$
\item \textbf{LZ2}: $\ell_i^{(2)} = q_i - 2q_{i-1} + q_{i-2}$, \quad $i \ge 2$
\item \textbf{LZ1 + centering}: apply LZ1 to $\hat{q}_i = q_i - \mu$, where $\mu = \mathrm{round}\!\left(\bar{d} / (2 \cdot eb)\right)$ and $\bar{d} = \frac{1}{256}\sum_{j} d_j$ is the tile's floating-point mean
\item \textbf{LZ2 + centering}: apply LZ2 to $\hat{q}_i = q_i - \mu$
\end{enumerate}

Adaptive centering subtracts the quantized tile mean $\mu$ before prediction. The mean is computed from the floating-point sum accumulated during the prediction loop and quantized once, avoiding per-element overhead. As illustrated in Figure~\ref{fig:centering}, this is most effective when the data has a large constant offset (e.g., pressure at 1000\,hPa): without centering, the boundary residual $q_0 = 1000$ dominates as the maximum absolute residual, forcing $r{=}10$ for the entire block (\step{a}). With centering, this residual drops to $q_0 - \mu = -1$, so the maximum becomes 3 and the rate falls to $r{=}2$, reducing the block cost from 44\,B to 16\,B (\step{b}). Centering is enabled only when its 4-byte cost for $\mu$ is outweighed by the savings.

\textbf{Error-bound preservation.} Centering does not affect the error bound. Quantization is the only lossy step: $q_i = \mathrm{round}(d_i / (2 \cdot eb))$ guarantees $|d_i - 2 \cdot eb \cdot q_i| \le eb$. All subsequent steps (centering, prediction, and encoding) are exact, invertible integer transforms of $q_i$. In particular, because $\mu$ is an integer in the quantized domain, decompression recovers $\hat{q}_i + \mu = q_i$ exactly for any $\mu$, so the reconstruction is identical with and without centering. Hence every prediction variant satisfies the same pointwise bound $|d_i - d_i'| \le eb$, and the four-way selection is purely a compression-ratio decision.

Each tile is divided into 8 blocks of 32 elements. For block $b$, let $r_b = \lceil \log_2(\max_{j \in \text{block}\, b} |\ell_j| + 1) \rceil$ denote the \emph{rate} (bits per element). The compressed size of block $b$ is:
\begin{equation}
S_b = \begin{cases} 0 & \text{if } r_b = 0 \\ 4 + 4 \cdot r_b & \text{otherwise} \end{cases}
\label{eq:block_cost}
\end{equation}
where the 4-byte sign map and $r_b$ bit-planes of 4 bytes each account for the fixed-length encoding. The total tile cost is $C = \sum_{b=0}^{7} S_b$ (plus 4 bytes for the mean if centering is enabled). {\fsz} selects the variant with the minimum $C$. The selection result is encoded in 2 bits of the first block's rate byte (bit 7 for LZ order, bit 6 for centering flag), requiring zero additional metadata.

\subsubsection{Single-Pass Four-Way Evaluation}
\label{sec:four_way_eval}

A naive implementation would require four separate passes over the tile data to evaluate each variant. {\fsz} collapses all four evaluations into a single pass by exploiting a key property of finite differences.

\textbf{Property.} \emph{For $k$-th order Lorenzo prediction, any constant offset $\mu$ cancels exactly for $i \ge k$:}
\begin{equation}
\Delta^k(q_i - \mu) = \Delta^k q_i, \quad \forall\, i \ge k
\label{eq:cancellation}
\end{equation}
\emph{where $\Delta^1 q_i = q_i - q_{i-1}$ and $\Delta^2 q_i = q_i - 2q_{i-1} + q_{i-2}$.}

\textbf{Proof.} The $k$-th order difference operator $\Delta^k$ is linear and $\Delta^k c = 0$ for any constant $c$ when $k \ge 1$. Hence $\Delta^k(q_i - \mu) = \Delta^k q_i - \Delta^k \mu = \Delta^k q_i$. \hfill $\square$

\textbf{Consequence.} The LZ1 and LZ2 residuals with centering are identical to the no-centering residuals for all elements except the first 1--2 boundary elements per tile. This means the per-block rates $r_b$ for blocks $b \ge 1$ are the same regardless of whether centering is applied. Only block 0's rate may change, and its corrected rate under centering can be computed from $\mu$ with a trivial adjustment after the main loop.

This enables {\fsz} to compute both LZ1 and LZ2 residuals simultaneously in a single loop, accumulate the floating-point sum for mean computation, and derive all four variant costs with minimal additional arithmetic, all from a single data read of the tile data.

\subsubsection{GPU-Centric Optimizations}

The tile-based architecture enables several GPU-specific optimizations that are critical for maintaining high throughput:
\begin{itemize}[leftmargin=10pt]
\item \textbf{No warp divergence.} Since the prediction order is selected \emph{per tile} and each of the 32 threads in a warp processes a different tile, all threads independently select their own tile's variant with no divergent branches within a warp.
\item \textbf{Vectorized data access.} The 256-element tile aligns naturally with \texttt{float4} loads (128-bit), processing 4 elements per memory transaction at peak bandwidth.
\item \textbf{Register-resident state.} The single-thread-per-tile model keeps all predictor states (LZ1, LZ2, mean accumulator, per-block maxima) in registers, avoiding shared memory or additional global memory accesses.
\item \textbf{PTX-level quantization.} The quantization fuses rounding and sign correction into a single PTX instruction sequence, eliminating branch overhead in the inner loop.
\end{itemize}

\aptLtoX[graphic=no,type=html]{\begin{shaded}\textbf{Takeaway II}: Naively evaluating four variants requires four data passes, quadrupling memory traffic. We collapse this into one pass because constant offsets cancel in finite differences for all but the first 1--2 elements, hiding the extra arithmetic behind the memory-bandwidth bottleneck.\end{shaded}}{\begin{columnshaded}
\vspace{3pt}
\textbf{Takeaway II}: Naively evaluating four variants requires four data passes, quadrupling memory traffic. We collapse this into one pass because constant offsets cancel in finite differences for all but the first 1--2 elements, hiding the extra arithmetic behind the memory-bandwidth bottleneck.
\end{columnshaded}}


\subsection{Fixed-Length Encoding and Bit-Shuffle}
\label{sec:encoding}

Each block of 32 residuals is encoded using fixed-length encoding: the maximum absolute residual determines the \emph{rate} (number of bits per element), and all 32 residuals are stored with exactly that many bits each. A 4-byte sign map records the sign of each element (one bit per element).

The encoded bits are written in \emph{bit-shuffled} order: bit-plane $b$ of all 32 elements is packed into a single \texttt{uchar4} (4 bytes), with element $i$'s bit $b$ stored at position $31 - i$ within the 32-bit word. This layout ensures that each bit-plane write is a single 4-byte aligned store, enabling efficient coalesced memory writes and avoiding irregular bit-shifting computations that would degrade GPU throughput.

\subsection{Global Synchronization via Decoupled Lookback}
\label{sec:sync}

Since each thread's compressed output has variable length, {\fsz} must compute cumulative byte offsets before writing output, a classic exclusive prefix-sum problem.

{\fsz} implements a three-level hierarchical prefix-sum:
\begin{enumerate}[leftmargin=15pt]
\item \textbf{Thread level:} Each thread sums its own compressed block sizes across all tiles.
\item \textbf{Warp level:} An inclusive prefix-sum is computed within each warp using \texttt{\_\_shfl\_up\_sync}, requiring $\log_2(32) = 5$ shuffle iterations.
\item \textbf{Device level:} The decoupled lookback technique~\cite{MerrillDecoupledLookback2016} computes the global prefix-sum across warps without a global barrier. Warps publish their local sums and scan backward through preceding warps to accumulate the prefix, using a three-state flag protocol (not started, local sum available, global prefix available) to coordinate without blocking.
\end{enumerate}

This keeps the entire prefix-sum on-GPU within the encoding kernel, preserving {\fsz}'s single-kernel architecture.

\aptLtoX[graphic=no,type=html]{\begin{shaded}\textbf{Takeaway III}: Single-kernel variable-length encoding imposes a fundamental two-pass constraint. {\fsz} turns this into an advantage: it fills the bandwidth-bound first read's idle compute with cross-block prediction and four-way evaluation, gaining richer prediction at no extra bandwidth.\end{shaded}}{\begin{columnshaded}
\vspace{3pt}
\textbf{Takeaway III}: Single-kernel variable-length encoding imposes a fundamental two-pass constraint. {\fsz} turns this into an advantage: it fills the bandwidth-bound first read's idle compute with cross-block prediction and four-way evaluation, gaining richer prediction at no extra bandwidth.
\end{columnshaded}}

\section{Experimental Evaluation}
\label{sec:evaluation}

We evaluate {\fsz} against state-of-the-art fast GPU lossy compressors from three perspectives: compression ratio, throughput, and data quality.

\subsection{Experimental Setup}
\label{sec:experimental-setup}

\textbf{Platform.} We evaluate all compressors on one NVIDIA GH200 GPU (96~GB HBM3, 132 SMs, SM~9.0) with CUDA Toolkit 13.0. Throughput is measured end-to-end (input on GPU to compressed output on GPU, and vice versa), averaged over 10 runs after warmup.

\textbf{Datasets and error bounds.} We use 8 real-world HPC datasets (84 fields) from SDRBench~\cite{Zhao2020SDRBench} and Open-SciVis~\cite{SciVis} (Table~\ref{tab:datasets}), evaluated at 3 value-range-based relative error bounds (REL): 1E-2, 1E-3, and 1E-4. A relative bound REL corresponds to the per-field absolute bound $eb = \text{REL} \times (d_{\max} - d_{\min})$, and every reconstructed value satisfies the pointwise guarantee $|d_i - d_i'| \le eb$.

\begin{table}[ht]
    \centering
    \FloatBodyStyle\footnotesize
    \renewcommand{\arraystretch}{1.05}
    \begin{tabular}{l r r l}
        \toprule
        \textbf{Dataset} & \textbf{Fields} & \textbf{Size (MB)} & \textbf{Domain} \\
        \midrule
        \rowcolor{gray!10} CESM-ATM~\cite{cesm} & 33 & 21,209 & Climate simulation \\
        EXAALT~\cite{exaalt2018} & 6 & 66 & Molecular dynamics \\
        \rowcolor{gray!10} HACC~\cite{hacc2016} & 6 & 6,431 & Cosmology simulation \\
        Hurricane~\cite{hurricane} & 13 & 1,240 & Weather simulation \\
        \rowcolor{gray!10} Miranda~\cite{miranda2005} & 7 & 1,008 & Turbulence simulation \\
        NYX~\cite{nyx} & 6 & 3,072 & Cosmology simulation \\
        \rowcolor{gray!10} Truss~\cite{synthetic_truss_with_five_defects} & 1 & 6,592 & CT scan simulation \\
        SCALE~\cite{scaleletkf2015} & 12 & 6,460 & Weather simulation \\
        \bottomrule
    \end{tabular}
    \vspace{3pt}
    \caption{Real-world HPC datasets used in our evaluation. All datasets are publicly available online: Truss from Open-SciVis~\cite{SciVis}; all others from SDRBench~\cite{Zhao2020SDRBench}.}
    \label{tab:datasets}
\end{table}

\textbf{Baselines.} We compare against {\cuszpP} and {\cuszpO}~\cite{cuSZp2023, Huang2024cuSZp2, Huang2025cuSZp3}, {\fzgpu}~\cite{ZhangFZ-GPU2023}, and cuZFP~\cite{cuzfp} (fixed-rate at 4, 8, and 16 bits per value). CPU-GPU hybrid compressors (e.g., cuSZ, MGARD-GPU) are excluded because their reliance on CPU-side computation limits end-to-end throughput to an order of magnitude below pure-GPU compressors~\cite{cuSZp2023}.

\subsection{Compression Ratio}
\label{sec:eval-cr}

Table~\ref{tab:cr_comparison} presents the compression ratio per dataset at three error bounds: each cell reports the min$\sim$max range and the average $\pm$ standard deviation across the dataset's fields. Rows are ordered so that {\fsz} is directly followed by {\cuszpO}, the strongest baseline in compression ratio. The highest average per dataset and error bound is highlighted in \textbf{bold}.

\begin{table*}[t]
    \centering
    \FloatBodyStyle\scriptsize
    \renewcommand{\arraystretch}{0.90}
    \setlength{\tabcolsep}{2pt}
    \begin{tabular}{c c cccccccc}
        \toprule
        & \textbf{REL} & \textbf{CESM-ATM} & \textbf{EXAALT} & \textbf{HACC} & \textbf{Hurricane} & \textbf{Miranda} & \textbf{NYX} & \textbf{Truss} & \textbf{SCALE} \\
        \midrule
        \multirow{6}{*}{\textbf{{\fsz}}}
        & \multirow{2}{*}{1E-2} & 22.49$\sim$92.59 & 5.51$\sim$6.90 & 11.41$\sim$51.96 & 13.94$\sim$89.52 & 28.72$\sim$47.67 & 15.12$\sim$127.80 & 13.20$\sim$13.20 & 20.84$\sim$109.32 \\
        &  & \BestCr{54.25$\pm$16.40} & \BestCr{6.07$\pm$0.64} & \BestCr{24.83$\pm$16.76} & \BestCr{43.40$\pm$24.45} & \BestCr{37.60$\pm$8.34} & \BestCr{70.52$\pm$56.81} & \BestCr{13.20} & \BestCr{51.68$\pm$32.78} \\
        \cmidrule{2-10}
        & \multirow{2}{*}{1E-3} & 14.05$\sim$59.97 & 3.51$\sim$3.98 & 5.67$\sim$14.07 & 8.68$\sim$58.38 & 14.17$\sim$32.78 & 10.39$\sim$125.50 & 6.54$\sim$6.54 & 11.51$\sim$80.22 \\
        &  & \BestCr{27.18$\pm$10.91} & \BestCr{3.72$\pm$0.23} & \BestCr{8.75$\pm$3.73} & 24.99$\pm$16.58 & \BestCr{23.00$\pm$7.89} & \BestCr{42.47$\pm$47.40} & \BestCr{6.54} & \BestCr{30.53$\pm$23.16} \\
        \cmidrule{2-10}
        & \multirow{2}{*}{1E-4} & 8.56$\sim$39.88 & 2.59$\sim$2.84 & 3.61$\sim$6.45 & 4.88$\sim$37.67 & 8.32$\sim$20.53 & 5.91$\sim$97.93 & 4.28$\sim$4.28 & 6.62$\sim$50.94 \\
        &  & \BestCr{16.09$\pm$6.83} & \BestCr{2.70$\pm$0.12} & \BestCr{4.68$\pm$1.28} & \BestCr{15.82$\pm$11.07} & \BestCr{14.01$\pm$5.32} & \BestCr{24.30$\pm$36.30} & \BestCr{4.28} & \BestCr{18.46$\pm$14.51} \\
        \midrule
        \multirow{6}{*}{\textbf{{\cuszpO}}}
        & \multirow{2}{*}{1E-2} & 18.44$\sim$82.38 & 5.51$\sim$6.82 & 10.94$\sim$19.90 & 13.26$\sim$89.69 & 16.18$\sim$47.35 & 14.36$\sim$127.80 & 12.97$\sim$12.97 & 16.80$\sim$109.55 \\
        &  & 42.98$\pm$18.62 & 6.06$\pm$0.62 & 14.68$\pm$4.17 & 42.16$\pm$25.81 & 34.44$\pm$11.69 & 69.14$\pm$58.40 & 12.97 & 46.19$\pm$35.19 \\
        \cmidrule{2-10}
        & \multirow{2}{*}{1E-3} & 12.99$\sim$57.43 & 3.51$\sim$3.96 & 5.65$\sim$11.84 & 8.68$\sim$58.50 & 11.78$\sim$29.24 & 10.50$\sim$125.56 & 6.47$\sim$6.47 & 11.10$\sim$79.70 \\
        &  & 24.53$\pm$10.88 & 3.72$\pm$0.22 & 8.06$\pm$2.86 & \BestCr{25.00$\pm$16.66} & 19.10$\pm$7.96 & 41.75$\pm$47.97 & 6.47 & 29.52$\pm$23.10 \\
        \cmidrule{2-10}
        & \multirow{2}{*}{1E-4} & 7.85$\sim$39.00 & 2.59$\sim$2.84 & 3.59$\sim$6.04 & 4.81$\sim$37.71 & 7.13$\sim$17.75 & 5.43$\sim$98.37 & 4.25$\sim$4.25 & 6.31$\sim$49.95 \\
        &  & 14.98$\pm$6.55 & 2.69$\pm$0.11 & 4.52$\pm$1.13 & 15.70$\pm$11.21 & 11.58$\pm$4.75 & 24.12$\pm$36.64 & 4.25 & 17.92$\pm$14.37 \\
        \midrule
        \multirow{6}{*}{\textbf{{\cuszpP}}}
        & \multirow{2}{*}{1E-2} & 3.88$\sim$69.43 & 5.51$\sim$6.72 & 5.24$\sim$10.08 & 5.94$\sim$88.87 & 4.44$\sim$44.71 & 9.60$\sim$127.80 & 12.67$\sim$12.67 & 3.88$\sim$105.89 \\
        &  & 32.56$\pm$20.23 & 5.95$\pm$0.49 & 7.63$\pm$2.59 & 38.70$\pm$28.32 & 29.59$\pm$14.61 & 66.73$\pm$61.01 & 12.67 & 37.76$\pm$37.27 \\
        \cmidrule{2-10}
        & \multirow{2}{*}{1E-3} & 2.78$\sim$39.01 & 3.51$\sim$3.94 & 3.43$\sim$5.20 & 3.71$\sim$56.88 & 3.12$\sim$25.96 & 5.09$\sim$125.55 & 6.37$\sim$6.37 & 2.75$\sim$72.60 \\
        &  & 14.53$\pm$9.33 & 3.69$\pm$0.19 & 4.31$\pm$0.94 & 22.31$\pm$18.05 & 15.40$\pm$8.96 & 38.44$\pm$50.42 & 6.37 & 21.11$\pm$22.75 \\
        \cmidrule{2-10}
        & \multirow{2}{*}{1E-4} & 2.11$\sim$24.55 & 2.59$\sim$2.83 & 2.53$\sim$3.39 & 2.70$\sim$36.66 & 2.33$\sim$16.25 & 3.35$\sim$98.23 & 4.21$\sim$4.21 & 2.14$\sim$42.06 \\
        &  & 8.26$\pm$5.26 & 2.68$\pm$0.10 & 2.96$\pm$0.47 & 14.36$\pm$11.79 & 9.17$\pm$5.74 & 22.14$\pm$37.65 & 4.21 & 12.34$\pm$12.83 \\
        \midrule
        \multirow{6}{*}{\textbf{{\fzgpu}}}
        & \multirow{2}{*}{1E-2} & 3.33$\sim$3.85 & \multirow{2}{*}{\scriptsize\shortstack{N.A.\\(bug)}} & 2.83$\sim$3.26 & 3.00$\sim$3.80 & 3.43$\sim$3.61 & 3.08$\sim$3.93 & \multirow{2}{*}{\scriptsize\shortstack{N.A.\\(bug)}} & 3.22$\sim$3.88 \\
        &  & 3.64$\pm$0.13 &  & 2.99$\pm$0.18 & 3.41$\pm$0.28 & 3.54$\pm$0.08 & 3.50$\pm$0.43 &  & 3.51$\pm$0.26 \\
        \cmidrule{2-10}
        & \multirow{2}{*}{1E-3} & 3.09$\sim$3.76 & \multirow{2}{*}{\scriptsize\shortstack{N.A.\\(bug)}} & 2.25$\sim$2.57 & 2.78$\sim$3.65 & 3.07$\sim$3.51 & 2.84$\sim$3.91 & \multirow{2}{*}{\scriptsize\shortstack{N.A.\\(bug)}} & 2.93$\sim$3.78 \\
        &  & 3.38$\pm$0.18 &  & 2.37$\pm$0.14 & 3.13$\pm$0.32 & 3.32$\pm$0.20 & 3.18$\pm$0.43 &  & 3.26$\pm$0.34 \\
        \cmidrule{2-10}
        & \multirow{2}{*}{1E-4} & 2.62$\sim$3.62 & \multirow{2}{*}{\scriptsize\shortstack{N.A.\\(bug)}} & 1.83$\sim$2.09 & 2.36$\sim$3.43 & 2.71$\sim$3.26 & 2.40$\sim$3.75 & \multirow{2}{*}{\scriptsize\shortstack{N.A.\\(bug)}} & 2.58$\sim$3.63 \\
        &  & 3.07$\pm$0.23 &  & 1.93$\pm$0.12 & 2.78$\pm$0.37 & 3.01$\pm$0.26 & 2.77$\pm$0.50 &  & 3.03$\pm$0.38 \\
        \bottomrule
    \end{tabular}
    \vspace{3pt}
    \caption{Compression ratio of 4 GPU lossy compressors. Each cell: min$\sim$max over the dataset's fields, with average$\pm$standard deviation below; single-field datasets omit the deviation. Best averages in \textbf{bold blue}.}
    \label{tab:cr_comparison}
\end{table*}

{\fsz} achieves the highest compression ratio across all error-bounded compressors on every dataset at every error bound except one case (Hurricane at REL~1E-3, where {\cuszpO} leads by 0.01). The advantage is most pronounced on fields whose data patterns align with {\fsz}'s richer prediction. At REL~1E-2, {\fsz} achieves 2.92$\times$ higher CR than {\cuszpO} on CESM-ATM Z3 (geopotential height with a smooth vertical gradient that second-order prediction captures), 2.61$\times$ on HACC yy (particle y-position with a near-linear distribution), 2.20$\times$ on CESM-ATM T (temperature in Kelvin with a large constant offset that centering removes), and 2.00$\times$ on Miranda density (turbulence density with strong spatial structure that cross-block prediction exploits). The per-field analysis in Section~\ref{sec:eval-perfield} provides further detail.
Even on datasets where {\cuszpO} is already strong (e.g., NYX), {\fsz} still improves CR by 4--8\% in geometric mean. Compared with {\fzgpu}, {\fsz} achieves 5--12$\times$ higher CR (geometric mean) because {\fsz}'s richer prediction leaves far smaller residuals to encode than {\fzgpu}'s. cuZFP is evaluated separately via rate-distortion (Section~\ref{sec:eval-quality}) since its fixed-rate mode does not correspond to error-bounded CR.

\aptLtoX[graphic=no,type=html]{\begin{shaded}\textbf{Observation I}: {\fsz} beats {\cuszpO} by up to 2.92$\times$ and {\cuszpP} by up to 10.95$\times$ in CR, capturing patterns a fixed first-order predictor misses through adaptive multi-order prediction, centering, and cross-block state.\end{shaded}}{\begin{columnshaded}
\vspace{3pt}
\textbf{Observation I}: {\fsz} beats {\cuszpO} by up to 2.92$\times$ and {\cuszpP} by up to 10.95$\times$ in CR, capturing patterns a fixed first-order predictor misses through adaptive multi-order prediction, centering, and cross-block state.
\end{columnshaded}}

\subsection{Throughput}
\label{sec:eval-throughput}

\begin{figure*}[t]
    \centering
    \includegraphics[width=\linewidth]{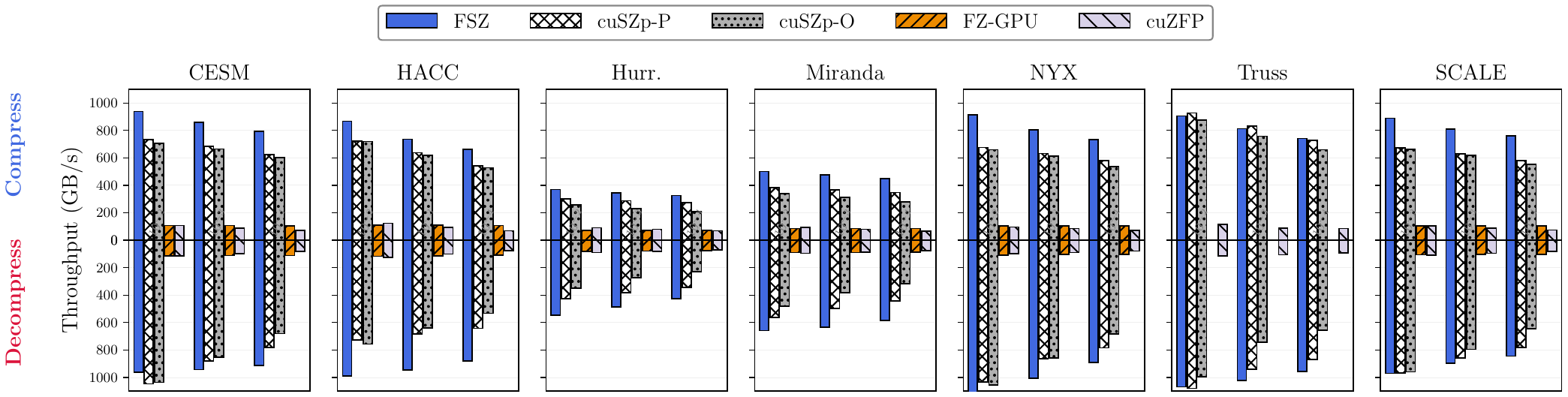}
    \caption{Compression (top) and decompression (bottom, mirrored) throughput at three error bounds per dataset. {\fsz} (solid blue) consistently achieves the highest throughput.}
    \label{fig:throughput_mirror}
\end{figure*}

\begin{figure}[t]
    \centering
    \includegraphics[width=\columnwidth]{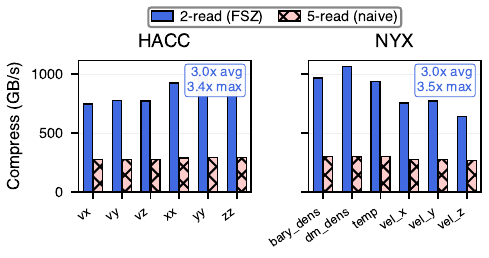}
    \caption{Single-pass four-way evaluation (blue) vs.\ naive 5-read baseline (gray) on HACC and NYX at REL~1E-2. Both produce identical CR; the speedup averages 2.8$\times$ across all 84 fields and three error bounds.}
    \label{fig:pass_study}
\end{figure}

Section~\ref{sec:rethinking} argued that the rate pass has arithmetic headroom for richer prediction. Figure~\ref{fig:throughput_mirror} confirms this empirically: despite performing $4\times$ richer prediction, {\fsz} achieves the highest throughput among all evaluated compressors in nearly every case. This is due to three complementary factors: (1)~the single-pass four-way evaluation that evaluates all prediction variants from a single data read without additional bandwidth consumption; (2)~GPU-centric optimizations including vectorized \texttt{float4}/\texttt{uchar4} access, PTX-level quantization, register-resident prediction state, and warp-divergence-free adaptive selection; and (3)~better prediction producing lower per-block rates, reducing encoding and decoding work.

\textbf{Compression throughput.} {\fsz} averages 676~GB/s, outperforming {\cuszpP} (536~GB/s) by 1.26$\times$, {\cuszpO} (507~GB/s) by 1.33$\times$, cuZFP (68--99~GB/s) by 7--10$\times$, and {\fzgpu} (98~GB/s) by 6--7$\times$. At REL~1E-4, the per-field speedup over {\cuszpO} reaches 1.65$\times$ on Hurricane QRAIN, 1.64$\times$ on Miranda diffusivity, and 1.63$\times$ on Hurricane QGRAUP.

\textbf{Decompression throughput.} {\fsz} averages 785~GB/s, outperforming {\cuszpP} (709~GB/s) by 1.11$\times$ and {\cuszpO} (648~GB/s) by 1.21$\times$ on average. At REL~1E-4, the per-field speedup over {\cuszpO} reaches 2.31$\times$ on Hurricane W, 2.09$\times$ on NYX velocity\_z, and 2.07$\times$ on Miranda velocity\_x.

\textbf{Single-pass evaluation benefit.} To quantify the impact of the single-pass four-way evaluation, we compare {\fsz}'s 2-read design against a naive 5-read baseline that evaluates each variant in a separate data pass. Both produce \emph{identical} compression ratios. The single-pass design achieves on average 2.8$\times$ higher compression throughput across all fields and error bounds (Figure~\ref{fig:pass_study}), confirming that the cancellation property eliminates what would otherwise be a major bandwidth penalty.

\aptLtoX[graphic=no,type=html]{\begin{shaded}\textbf{Observation II}: Better prediction lowers per-block rates, cutting encode/decode work and raising CR and throughput. With GPU optimizations, {\fsz} reaches 676/785~GB/s (de)compression, the highest of all compressors.\end{shaded}}{\begin{columnshaded}
\vspace{3pt}
\textbf{Observation II}: Better prediction lowers per-block rates, cutting encode/decode work and raising CR and throughput. With GPU optimizations, {\fsz} reaches 676/785~GB/s (de)compression, the highest of all compressors.
\end{columnshaded}}

\subsection{Data Quality}
\label{sec:eval-quality}

\begin{figure*}[t]
    \centering
    \includegraphics[width=\linewidth]{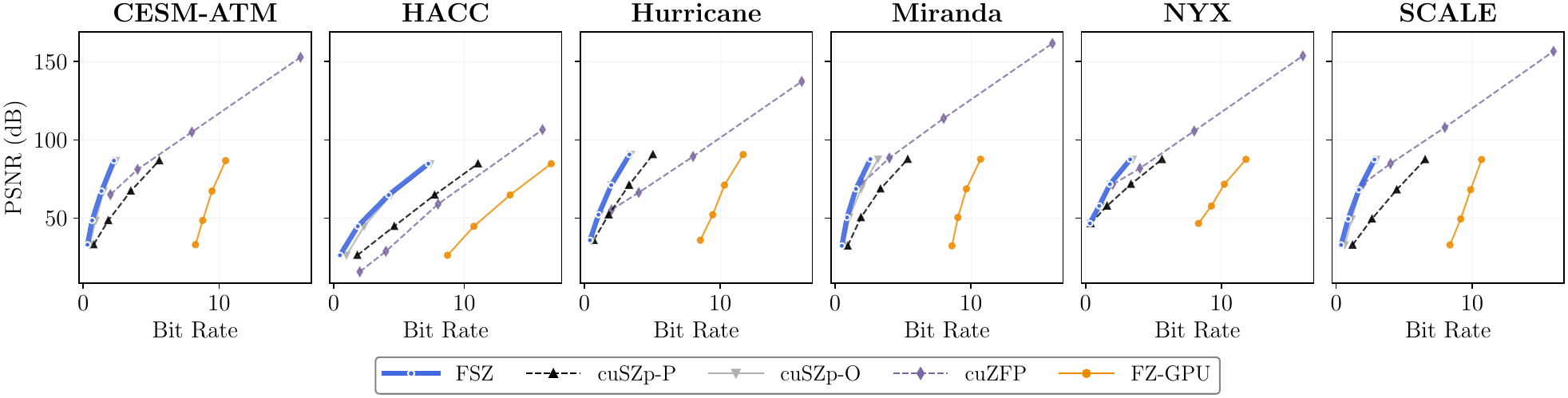}
    \caption{Rate-distortion curves (PSNR vs bit rate) for {\fsz} and baseline compressors on 6 representative datasets. Higher and further left is better. {\fsz} consistently achieves the best rate-distortion trade-off.}
    \label{fig:rate_distortion}
\end{figure*}

\begin{figure*}[t]
    \centering
    \includegraphics[width=\textwidth]{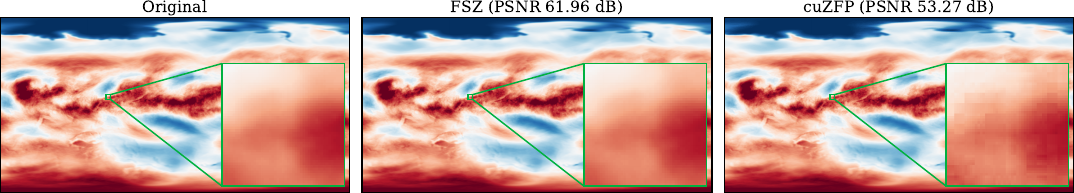}
    \caption{Visual quality on CESM-ATM RELHUM at matched CR ({\fsz} 14.88, cuZFP 14.86). Each panel shows the full slice with a $48\times48$ zoom inset. {\fsz} is indistinguishable from the original; cuZFP breaks the fine-scale structure into $4\times4$ blocks. Field-wide, {\fsz} holds its error bound $eb$ at every point, while cuZFP exceeds it at 23.0\% of points.}
    \label{fig:vis_quality}
\end{figure*}

We evaluate data quality using rate-distortion analysis and visualization. Figure~\ref{fig:rate_distortion} presents rate-distortion curves (PSNR vs bit rate) for all compressors. A curve that is higher and further left indicates better quality at higher compression. {\fsz} consistently achieves the best rate-distortion performance among all error-bounded compressors, matching or exceeding {\cuszpO} at every bit rate across all six datasets. cuZFP generally achieves lower PSNR than {\fsz} at comparable bit rates because its fixed-rate mode ignores data characteristics.

Since {\fsz}, {\cuszpP}, {\cuszpO}, and {\fzgpu} share the same quantization strategy, they produce identical reconstructed data at the same error bound, so their reconstructions are indistinguishable. We therefore compare {\fsz} against cuZFP, which uses a fundamentally different fixed-rate encoding. Figure~\ref{fig:vis_quality} visualizes a slice of CESM-ATM RELHUM at matched compression ratio, each panel showing the full slice with a $48\times48$ zoom inset. cuZFP is configured at fixed-rate 2, which it stores at an effective 2.154 bits per value because cuZFP pads the 26-level dimension of the field up to 28, giving CR 14.86 against {\fsz}'s 14.88. {\fsz}'s reconstruction is visually identical to the original data (PSNR 61.96~dB), while cuZFP breaks the fine-scale structure into $4\times4$ blocks (PSNR 53.27~dB). The same separation holds pointwise across the whole domain, not only in the inset. Over the full field of 168.5M points, {\fsz} holds its error bound at every point, while cuZFP, whose fixed-rate mode carries no pointwise guarantee, exceeds that same bound at 23.0\% of points.

\aptLtoX[graphic=no,type=html]{\begin{shaded}\textbf{Observation III}: {\fsz} achieves the best rate-distortion trade-off and, at matched CR, delivers 8.7~dB higher PSNR than cuZFP with no visible block artifacts. {\fsz} holds its error bound at every point of the field, while cuZFP exceeds that same bound at 23.0\% of points.\end{shaded}}{\begin{columnshaded}
\vspace{3pt}
\textbf{Observation III}: {\fsz} achieves the best rate-distortion trade-off and, at matched CR, delivers 8.7~dB higher PSNR than cuZFP with no visible block artifacts. {\fsz} holds its error bound at every point of the field, while cuZFP exceeds that same bound at 23.0\% of points.
\end{columnshaded}}

\section{Further Analysis}
\label{sec:eval-further}

\subsection{Per-Field Analysis}
\label{sec:eval-perfield}

\begin{figure*}[t]
    \centering
    \includegraphics[width=\linewidth]{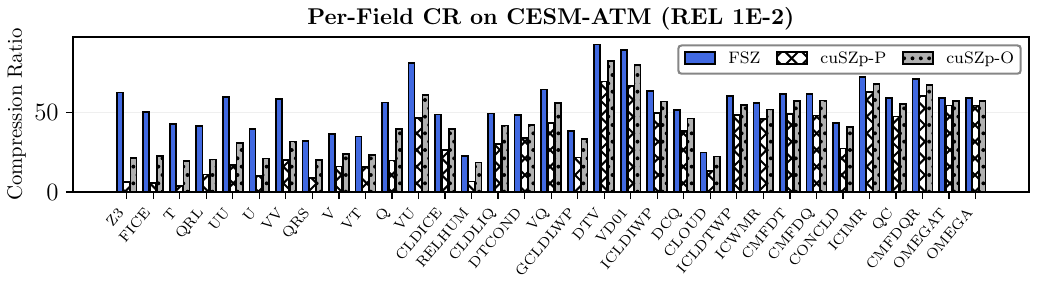}
    \caption{Per-field CR comparison on all 33 CESM-ATM fields at REL~1E-2, sorted by {\fsz}'s advantage over {\cuszpO}. {\fsz} achieves up to 2.92$\times$ higher CR on fields with strong gradients (Z3, T, FICE).}
    \label{fig:perfield_cesm}
\end{figure*}

To provide deeper insight into where {\fsz}'s adaptive prediction is most effective, Figure~\ref{fig:perfield_cesm} presents per-field compression ratio comparisons on all 33 CESM-ATM fields at REL~1E-2, sorted by {\fsz}/{\cuszpO} ratio.

The fields where {\fsz} achieves the largest improvements (2--2.92$\times$) share a common characteristic: they exhibit strong linear or polynomial trends that second-order Lorenzo prediction captures effectively. For example, CESM-ATM Z3 (geopotential height) has a smooth vertical gradient, and HACC yy (particle position in y-direction) has a near-linear distribution. In contrast, Hurricane~W (vertical velocity) is a near-tie because it has highly localized turbulent patterns where neither LZ1 nor LZ2 provides a consistent advantage.

\aptLtoX[graphic=no,type=html]{\begin{shaded}\textbf{Observation IV}: {\fsz}'s largest per-field gains (2--2.92$\times$ over {\cuszpO}) fall on fields with strong linear or polynomial trends (e.g., CESM-ATM Z3, HACC yy), where adaptive multi-order prediction and centering excel.\end{shaded}}{\begin{columnshaded}
\vspace{3pt}
\textbf{Observation IV}: {\fsz}'s largest per-field gains (2--2.92$\times$ over {\cuszpO}) fall on fields with strong linear or polynomial trends (e.g., CESM-ATM Z3, HACC yy), where adaptive multi-order prediction and centering excel.
\end{columnshaded}}

\subsection{Ablation Study}
\label{sec:eval-ablation}

To validate that each of {\fsz}'s three designs contributes meaningfully and that their benefits compound, we conduct an ablation study by progressively enabling features. The baseline is a cuSZp-equivalent compressor: first-order Lorenzo prediction, no centering, and independent block processing. Table~\ref{tab:ablation} presents the per-field CR buildup on Miranda (7 fields) at three error bounds.

\begin{table}[t]
    \centering
    \FloatBodyStyle\footnotesize
    \renewcommand{\arraystretch}{1.05}
    \setlength{\tabcolsep}{4pt}
    \begin{tabular}{c l r r r r}
        \toprule
        \textbf{REL} & \textbf{Field} & \textbf{Base} & \textbf{+Cross} & \textbf{+Cent.} & \textbf{+LZ2} \\
        \midrule
        \multirow{7}{*}{1E-2}
        & \cellcolor{gray!10}density      & \cellcolor{gray!10} 4.44 & \cellcolor{gray!10}18.42 & \cellcolor{gray!10}28.59 & \cellcolor{gray!10}32.39 \\
        & diffusivity  & 44.64 & 47.02 & 47.02 & 47.61 \\
        & \cellcolor{gray!10}pressure     & \cellcolor{gray!10}40.02 & \cellcolor{gray!10}42.38 & \cellcolor{gray!10}42.39 & \cellcolor{gray!10}43.36 \\
        & velocity\_x  & 23.53 & 27.35 & 27.35 & 28.72 \\
        & \cellcolor{gray!10}velocity\_y  & \cellcolor{gray!10}23.56 & \cellcolor{gray!10}28.98 & \cellcolor{gray!10}28.99 & \cellcolor{gray!10}29.73 \\
        & velocity\_z  & 26.26 & 31.52 & 31.53 & 33.75 \\
        & \cellcolor{gray!10}viscosity    & \cellcolor{gray!10}44.71 & \cellcolor{gray!10}47.08 & \cellcolor{gray!10}47.08 & \cellcolor{gray!10}47.67 \\
        \midrule
        \multirow{7}{*}{1E-3}
        & \cellcolor{gray!10}density      & \cellcolor{gray!10} 3.12 & \cellcolor{gray!10}11.97 & \cellcolor{gray!10}18.06 & \cellcolor{gray!10}21.96 \\
        & diffusivity  & 25.91 & 28.81 & 28.83 & 32.73 \\
        & \cellcolor{gray!10}pressure     & \cellcolor{gray!10}21.27 & \cellcolor{gray!10}23.26 & \cellcolor{gray!10}23.26 & \cellcolor{gray!10}26.70 \\
        & velocity\_x  &  9.88 & 12.28 & 12.28 & 14.17 \\
        & \cellcolor{gray!10}velocity\_y  & \cellcolor{gray!10} 9.93 & \cellcolor{gray!10}13.40 & \cellcolor{gray!10}13.43 & \cellcolor{gray!10}15.43 \\
        & velocity\_z  & 11.71 & 14.76 & 14.78 & 17.21 \\
        & \cellcolor{gray!10}viscosity    & \cellcolor{gray!10}25.96 & \cellcolor{gray!10}28.87 & \cellcolor{gray!10}28.88 & \cellcolor{gray!10}32.78 \\
        \midrule
        \multirow{7}{*}{1E-4}
        & \cellcolor{gray!10}density      & \cellcolor{gray!10} 2.33 & \cellcolor{gray!10} 8.50 & \cellcolor{gray!10}12.60 & \cellcolor{gray!10}14.70 \\
        & diffusivity  & 16.23 & 17.55 & 17.56 & 20.50 \\
        & \cellcolor{gray!10}pressure     & \cellcolor{gray!10}12.66 & \cellcolor{gray!10}13.70 & \cellcolor{gray!10}13.70 & \cellcolor{gray!10}15.83 \\
        & velocity\_x  &  5.37 &  7.04 &  7.05 &  8.32 \\
        & \cellcolor{gray!10}velocity\_y  & \cellcolor{gray!10} 5.34 & \cellcolor{gray!10} 7.47 & \cellcolor{gray!10} 7.49 & \cellcolor{gray!10} 8.82 \\
        & velocity\_z  &  6.02 &  8.01 &  8.02 &  9.34 \\
        & \cellcolor{gray!10}viscosity    & \cellcolor{gray!10}16.25 & \cellcolor{gray!10}17.58 & \cellcolor{gray!10}17.59 & \cellcolor{gray!10}20.53 \\
        \bottomrule
    \end{tabular}
    \vspace{3pt}
    \caption{Ablation on Miranda: CR as features are progressively enabled. Base = LZ1 with independent blocks. +Cross = cross-block prediction state. +Cent.\ = adaptive centering. +LZ2 = adaptive multi-order prediction (= full {\fsz}).}
    \label{tab:ablation}
\end{table}

Each design carries the compression ratio on a different class of fields, and these classes do not overlap, so all three are required to cover the field set. First, \textbf{cross-block prediction state is most effective on fields with strong spatial structure.} On Miranda density, it alone improves CR from 4.44 to 18.42 (4.15$\times$) at REL~1E-2, because the density field has strong spatial structure where carrying prediction state across block boundaries eliminates large boundary residuals that would otherwise inflate the encoding rate of entire blocks. The effect is consistent across velocity fields (1.2--1.4$\times$) and on smooth fields like diffusivity (1.05--1.11$\times$) where prediction is already effective within single blocks. Across all 84 fields and 3 error bounds, cross-block prediction state provides a geometric mean CR improvement of 1.60$\times$.

Second, \textbf{adaptive centering is most effective on fields with large constant offsets.} On Miranda density, centering adds 1.48--1.55$\times$ on top of cross-block prediction state, and it stands aside on velocity and diffusivity fields, whose values already sit near zero and need no offset correction. Across all 84 fields, adaptive centering provides a geometric mean of 1.07$\times$, and it is the design that carries the offset-heavy fields: HACC positions gain up to 1.28$\times$, as do CESM-ATM temperature fields, where the other two designs have little offset to remove.

Third, \textbf{adaptive multi-order prediction is most effective at tight error bounds.} At REL~1E-2, second-order Lorenzo prediction (LZ2) contributes 1.00--1.05$\times$, since most blocks are already rate-0 regardless of prediction order. Once the bound tightens, the rate-0 shortcut no longer applies and curvature in the data governs the residual, so LZ2 contributes up to 1.22$\times$ (Miranda density at REL~1E-3, Table~\ref{tab:ablation}) and 1.15--1.18$\times$ on Miranda velocity and diffusivity fields at REL~1E-4. Across all 84 fields, its geometric mean improvement rises from 1.00$\times$ at REL~1E-2 to 1.05$\times$ at REL~1E-4, which is exactly the case the other two designs do not cover.

The three designs are therefore complementary rather than interchangeable: each delivers the gain on a class of fields or error bounds the other two do not reach, and none of them substitutes for another. Across all 84 fields, their combined geometric mean CR gain is 1.75$\times$ over the baseline.

\aptLtoX[graphic=no,type=html]{\begin{shaded}\textbf{Observation V}: The three designs are complementary, each covering a case the others do not: cross-block on structured fields (up to 4.15$\times$), centering on offset-heavy fields (up to 1.55$\times$), and multi-order at tight bounds (up to 1.22$\times$). Combined: 1.75$\times$ geometric mean.\end{shaded}}{\begin{columnshaded}
\vspace{3pt}
\textbf{Observation V}: The three designs are complementary, each covering a case the others do not: cross-block on structured fields (up to 4.15$\times$), centering on offset-heavy fields (up to 1.55$\times$), and multi-order at tight bounds (up to 1.22$\times$). Combined: 1.75$\times$ geometric mean.
\end{columnshaded}}

\subsection{Tile Configuration Analysis}
\label{sec:eval-tilesize}

\begin{figure}[!b]
    \centering
    \includegraphics[width=\columnwidth]{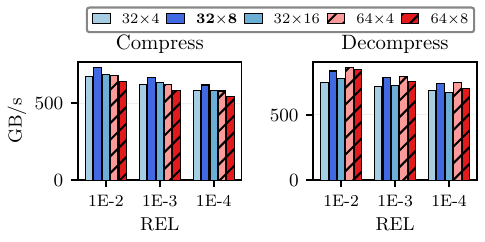}
    \caption{Throughput of 5 tile configurations at 3 error bounds. Default $B$=32/$T$=8 (dark blue) achieves the highest throughput.}
    \label{fig:tile_sweep}
\end{figure}

Recall that {\fsz} has two key configuration parameters: the \emph{block size} $B$ (elements per encoding block) and the \emph{number of blocks per tile} $T$. The tile size is $B \times T$ elements. Figure~\ref{fig:tile_sweep} visualizes the compression ratio and throughput of all 5 configurations at each error bound.

 A larger block reduces per-block metadata overhead (one rate byte per block), while a larger tile extends the cross-block prediction chain, reducing boundary residuals. However, both come with trade-offs: larger blocks make the fixed-length encoding less adaptive to local variations (a single outlier inflates the entire block's rate), and larger tiles increase register pressure and reduce GPU parallelism. To find the optimal balance, we sweep 5 configurations across 84 fields at 3 error bounds (252 tests per configuration), as shown in Table~\ref{tab:tile_sweep}.

\begin{table}[ht]
    \centering
    \FloatBodyStyle\footnotesize
    \renewcommand{\arraystretch}{1.05}
    \begin{tabular}{l r rrr r}
        \toprule
        \textbf{Config} & \textbf{Tile} & \textbf{Win} & \textbf{Tie} & \textbf{Loss} & \textbf{CR ratio} \\
        \midrule
        \rowcolor{gray!10} $B$=32, $T$=4 & 128 & 0 & 85 & 167 & 0.95$\times$ \\
        \textbf{$B$=32, $T$=8} & \textbf{256} & \multicolumn{3}{c}{\textit{baseline}} & \textbf{1.00$\times$} \\
        \rowcolor{gray!10} $B$=32, $T$=16 & 512 & 148 & 104 & 0 & 1.03$\times$ \\
        $B$=64, $T$=4 & 256 & 42 & 40 & 170 & 0.97$\times$ \\
        \rowcolor{gray!10} $B$=64, $T$=8 & 512 & 66 & 78 & 108 & 1.01$\times$ \\
        \bottomrule
    \end{tabular}
    \vspace{3pt}
    \caption{Tile configuration sweep: each row is compared against the default $B$=32, $T$=8. Win = $>$2\% higher CR. CR ratio is the average ratio to the baseline.}
    \label{tab:tile_sweep}
\end{table}

As shown in Table~\ref{tab:tile_sweep} and Figure~\ref{fig:tile_sweep}, increasing the number of blocks per tile (fixing $B$=32) consistently improves CR: $T$=16 \emph{never} loses to $T$=8, confirming that longer cross-block prediction chains are beneficial. However, doubling the block size to $B$=64 hurts at tight error bounds: $B$=64/$T$=4 loses 170 out of 252 tests because coarser blocks are more sensitive to outliers. $B$=64/$T$=8 partially recovers with a longer chain but still loses 108 tests and incurs 10--32\% throughput regression (Figure~\ref{fig:tile_sweep}). We select $B$=32, $T$=8 as the default: $B$=32 packs each bit-plane into exactly one 32-bit word so all encoded writes stay 4-byte aligned, and this configuration provides robust CR across all error bounds and the highest throughput among all configurations.

\aptLtoX[graphic=no,type=html]{\begin{shaded}\textbf{Observation VI}: Longer chains only help ($T$=16 never loses), while larger blocks hurt at tight bounds since one outlier inflates the block. Default $B$=32/$T$=8 gives robust CR, 4-byte-aligned encoding, and the highest throughput.\end{shaded}}{\begin{columnshaded}
\vspace{3pt}
\textbf{Observation VI}: Longer chains only help ($T$=16 never loses), while larger blocks hurt at tight bounds since one outlier inflates the block. Default $B$=32/$T$=8 gives robust CR, 4-byte-aligned encoding, and the highest throughput.
\end{columnshaded}}

\subsection{Portability: NVIDIA A100}
\label{sec:eval-a100}

\begin{figure}[t]
    \centering
    \includegraphics[width=\columnwidth]{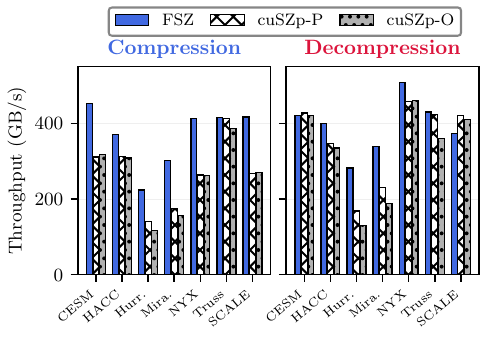}
    \caption{Compression and decompression throughput on NVIDIA A100 (80~GB) at REL~1E-3. {\fsz} outperforms both {\cuszp} modes on all datasets in compression, with larger speedups than on GH200.}
    \label{fig:a100_throughput}
\end{figure}

To verify that {\fsz}'s advantages are not specific to GH200, we repeat the throughput evaluation on NVIDIA A100 (80~GB HBM2e, SM~8.0) at REL~1E-3. As shown in Figure~\ref{fig:a100_throughput}, {\fsz} achieves 386~GB/s compression and 389~GB/s decompression, outperforming {\cuszpP} by 1.46$\times$ and {\cuszpO} by 1.49$\times$ in compression throughput (up to 1.93$\times$ on Hurricane), and up to 2.18$\times$ in decompression throughput (Hurricane vs {\cuszpO}). The speedup over {\cuszp} is \emph{larger} on A100 than on GH200 (1.46$\times$ vs 1.23$\times$ at the same error bound). This is because {\fsz}'s higher CR reduces the amount of data written during encoding, and this bandwidth saving is amplified on A100's lower-bandwidth memory. Additionally, {\fsz}'s single-pass four-way evaluation and GPU optimizations (vectorized access, PTX quantization, register-resident state, divergence-free selection) are effective across GPU generations.

\aptLtoX[graphic=no,type=html]{\begin{shaded}\textbf{Observation VII}: {\fsz}'s gains carry across GPU generations: its A100 speedup over {\cuszp} (1.46$\times$) exceeds GH200's (1.23$\times$) because higher CR writes fewer bytes, a saving amplified on lower-bandwidth memory.\end{shaded}}{\begin{columnshaded}
\vspace{3pt}
\textbf{Observation VII}: {\fsz}'s gains carry across GPU generations: its A100 speedup over {\cuszp} (1.46$\times$) exceeds GH200's (1.23$\times$) because higher CR writes fewer bytes, a saving amplified on lower-bandwidth memory.
\end{columnshaded}}

\section{Related Work}
\label{sec:related}

Error-bounded lossy compression for scientific data has been widely studied on CPUs, with notable compressors including SZ~\cite{Di2016SZ, Tao2017SZ, Liang2018SZ, Zhao2020SZauto, hybrid-sz} (prediction-based), ZFP~\cite{Lindstrom2014ZFP} (transform-based), and wavelet-based approaches~\cite{Huang-Exploring-Wavelet-Transform, Li2023SPERR}. These achieve high compression ratios but are inherently low throughput, motivating GPU-specific designs for throughput-sensitive applications including scientific simulations and large language model training~\cite{Huang2023RTM, WuLossyQuantumCircuit2019, Sasaki2015LossyCheckpointing, Huang2025ZCCL, huang2023ccoll, huang2023gzccl, Huang2025ghZCCL, touvron2023llama, Dai2024CommunicationLLM}. Early GPU compressors are limited in different ways: cuZFP~\cite{cuzfp} supports only fixed-rate compression on the GPU, while cuSZ~\cite{cusz2020} and cuSZx~\cite{Yu2022SZx} offload dependency-heavy stages to the CPU, limiting end-to-end throughput. The pure-GPU, single-kernel paradigm eliminates this bottleneck: cuSZp~\cite{cuSZp2023, Huang2024cuSZp2, Huang2025cuSZp3} fuses the entire pipeline into one kernel with fixed-length encoding, hierarchical prefix-sum, outlier encoding, and GPU-centric performance optimization; and FZ-GPU~\cite{ZhangFZ-GPU2023} replaces Huffman with bitshuffle encoding. These works have pushed throughput to hundreds of GB/s but all employ \emph{fixed first-order prediction}, because richer prediction was widely assumed to inevitably slow down the compressor due to additional data accesses and computation. In contrast, {\fsz} exploits the two-pass architecture's headroom to introduce cross-block state and adaptive multi-order prediction with centering \emph{without} extra data reads, breaking the prediction-throughput trade-off.

A separate line of recent GPU compressors instead maximizes compression ratio at the expense of throughput: cuSZ-Hi~\cite{cuszhi} couples interpolation-based prediction with an optimized lossless-encoding pipeline, and PFPL~\cite{pfpl} is a portable CPU/GPU error-bounded compressor. These reach higher compression ratios but run at substantially lower GPU throughput, outside the fast-compressor category that {\fsz} targets. Thus, we treat these high-ratio designs as complementary related work rather than direct throughput baselines.

\section{Conclusion and Future Work}
\label{sec:conclusion}

Prior fast GPU lossy compressors have treated prediction quality and GPU efficiency as fundamentally opposing goals, keeping prediction deliberately simple to preserve throughput. This paper shows that the tension is not irreconcilable. By co-designing the prediction stage with the GPU execution model, {\fsz} introduces three mutually reinforcing designs (cross-block prediction state, adaptive multi-order prediction with centering, and single-pass four-way evaluation) that align richer prediction with warp-level execution, register allocation, and memory access, all within a single CUDA kernel.

Experiments on NVIDIA GH200 with 8 scientific datasets (84 fields, 3 error bounds) confirm that {\fsz} outperforms {\cuszpO} by up to 2.92$\times$ and {\cuszpP} by up to 10.95$\times$ in compression ratio, while simultaneously achieving the highest throughput (676~GB/s compression, 785~GB/s decompression) among all evaluated GPU lossy compressors; similar speedups are confirmed on NVIDIA A100. These results demonstrate that the prediction stage is a rich source of compression ratio gains on GPUs when co-designed with the execution model. The current design is specific to GPUs. Looking ahead, we believe this co-design principle generalizes beyond GPUs and plan to extend it to other hardware platforms such as FPGAs and emerging AI accelerators (e.g., Cerebras WSE).

\section*{Acknowledgment}
This work was supported by startup funding from the University of South Florida.

\bibliographystyle{IEEEtran}
\bibliography{reference.bib}

\end{document}